\documentstyle[12pt, epsf]{article}
\def\ga{\mathrel{\raise.3ex\hbox{$>$\kern-.75em\lower1ex\hbox{$\sim$}}}}
\def\la{\mathrel{\raise.3ex\hbox{$<$\kern-.75em\lower1ex\hbox{$\sim$}}}}
\def\gev{{\rm \, Ge\kern-0.125em V}}
\def\tev{{\rm \, Te\kern-0.125em V}}
\def\beq{\begin{equation}}
\def\eeq{\end{equation}}
\def\ss{\scriptscriptstyle}
\def\mb{m_{\widetilde B}}
\def\msf{m_{\tilde f}}
\def\mtb{\overline{m}_{\ss t}}
\def\mbb{\overline{m}_{\ss b}}
\def\mfb{\overline{m}_{\ss f}}
\def\mf{m_{\ss{f}}}
\def\gt{\gamma_t}
\def\gb{\gamma_b}
\def\gf{\gamma_f}
\def\thm{\theta_\mu}

\begin{document}
\begin{titlepage}
\pagestyle{empty}
\baselineskip=21pt
\rightline{hep-ph/9502401}
\rightline{UMN--TH--1329/95}
\rightline{TPI--MINN--95/3}
\rightline{UCSBTH--95--3}
\rightline{February 1995}
\vskip1.25in
\begin{center}
{\large{\bf
Phases in the MSSM, Electric Dipole Moments and Cosmological
Dark Matter}}
\end{center}
\begin{center}
\vskip 0.5in
{Toby Falk,$^1$ Keith A.~Olive,$^1$  and Mark Srednicki$^2$
}\\
\vskip 0.25in
{\it
$^1${School of Physics and Astronomy,
University of Minnesota, Minneapolis, MN 55455, USA}\\
$^2${Department of Physics,
University of California, Santa Barbara, CA 93106, USA}\\}
\vskip 0.5in
{\bf Abstract}
\end{center}
\baselineskip=18pt \noindent
We consider the effect of CP violating phases in the MSSM on the relic density
of the lightest supersymmetric particle (LSP).  In particular, we find that the
upper limits on the LSP mass are relaxed when phases in the MSSM are
allowed to take non-zero values when the LSP is predominantly a gaugino
(bino).  Previous limits of $\mb \la 250$ GeV for $\Omega h^2 < 0.25$
can be relaxed to $\mb \la 650$ GeV.  We also consider the additional
constraints imposed by the neutron and electron electric dipole moments
induced by these phases. Though there is some restriction on the phases,
the bino mass may still be as large as $\sim$ 350 GeV and certain phases can
be arbitrarily large.
\end{titlepage}
\baselineskip=18pt
It is well known that by considering the cosmological relic density of
stable particles, one can establish mass limits on these particles.
For example, light neutrinos ($m_\nu < 1$ MeV) contribute too much to
the overall cosmological mass density if $m_\nu \ga 25$ eV for $\Omega
h^2 < 1/4$ \cite{cows}. The relic density of heavier neutrinos are
determined by the their annihilation cross section and yield lower
bounds to neutrino masses $m_\nu \ga$ 4--7 GeV \cite{lw,swo}. In the
minimal supersymmetric standard model (MSSM), because of the many
unknown parameters it becomes a significantly more complicated task to
set limits on the mass of the lightest particle, the LSP, e.g.
the annihilation cross section will depend on several parameters which
determine the identity of the LSP. Here we will consider only neutralinos as
the LSP (for a recent discussion on the possibilities for sneutrino dark
matter, see
\cite{fkosi}). The mass limits will then depend sensitively on the
masses of the scalars \cite{phot}, and in general, limits are obtained
on portions of the parameter space \cite{ehnos}. For example when
parameters are chosen so that the LSP is a gaugino (when the
supersymmetry breaking gaugino masses $M_{1,2}$ are taken to be
smaller than the supersymmetric Higgs mixing mass, $\mu$), for a given
set of scalar masses, the requirement that $\Omega h^2 < 1/4$ places a lower
bound on the gaugino mass. For small gaugino masses, this
corresponds to a lower limit on the photino mass \cite{phot}.  At larger
gaugino
masses (but still smaller than $\mu$), the LSP is a bino
\cite{osi3}. As the bino mass is increased, there is an upper limit
$\mb \la$ 250 GeV \cite{osi3,gkt}. In \cite{fkmos}, it was shown that
this upper limit is sensitive to the level of sfermion mixing.  Here, we
find that the upper limit to the bino mass is relaxed when phases in
the MSSM, and in particular the phases associated with the
off-diagonal sfermion masses, are allowed to take non-zero values. Now, the
upper limit is increased to $\mb \la 650$ GeV.

There has been a considerable amount of work concerning phases in the MSSM. For
the most part these phases are ignored because they tend to induce large
electric dipole moments for the neutron \cite{dn,dgh}. To suppress the electric
dipole moments, either large scalar masses (approaching $1 \tev$) or small
angles
(of order $10^{-3}$, when all SUSY masses are of order 100 GeV) are required.
For the most part, the community has opted for the latter, though the
possibilities for large phases was recently considered in \cite{ko}. To
reconcile large phases with small electric dipole moments, some of the
sparticle
masses are required to be heavy. In \cite{ko}, either large sfermion
or neutralino masses (or both) were required. However, unless
$R$-parity is broken and the LSP is not stable, one would require that the
sfermions be heavier than the neutralinos.   If they are much heavier, this
would result in an excessive relic density of neutralinos.  The object
of this letter is to determine the relationship between potentially large
phases
in the MSSM and the relic density while remaining consistent with experimental
bounds on the electric dipole moments.

In the MSSM, the possibility for new phases arises from a number of
sources.  First, in the superpotential, there is the Higgs mixing
mass, $\mu$.  There are also several parameters
associated with supersymmetry breaking: gaugino masses $M_i$, $i=1$--$3$
(we will assume GUT conditions on these masses so that we need
only consider one of them, $M_2$); and in the scalar sector, soft scalar
masses;
soft bilinear, $B\mu$, and trilinear, $A_f$, terms.  Not all of these phases
are physical \cite{dgh}. It is common to rotate away the phase of the
gaugino masses, and to make $B\mu$ real, which ensures that the vacuum
expectation values of the Higgs fields are real. We will also, for
simplicity, ignore generation mixing in the sfermion sector though we will
include
left-right mixing. If furthermore we assume that all of the $A_f$'s are
equal, and simply label them by $A$, we are left with two independent phases,
the phase of $A$, $\theta_A$, and the phase of $\mu$, $\theta_\mu$.
The phase of $B$ is fixed by the condition that $B\mu$ is real.

As both the relic cosmological density of neutralinos and the electric dipole
moments are strongly dependent on the sfermion masses, it will be very useful
to identify the combination of phases that enter the sfermion mass$^2$ matrix.
We take the general form of the sfermion mass$^2$ matrix to be  \cite{er}
\beq
\pmatrix{ M_L^2 + m_f^2 + \cos 2\beta (T_{3f} - Q_f\sin ^2 \theta_W) M_Z^2 &
m_f\,\overline{m}_{\ss f} e^{i \gamma_f}
\cr
\noalign{\medskip} m_f\,\overline{m}_{\ss f} e^{-i \gamma_f} & M_R^2 + m_f^2 +
\cos 2\beta Q_f\sin ^2
\theta_W M_Z^2
\cr }~
\eeq
where $M_{L(R)}$ are the soft supersymmetry breaking sfermion mass which we
have assumed are generation independent and generation diagonal and hence real.
We will also assume
 $M_L = M_R$.
$m_f$ is the mass of the fermion
$f$,
$\tan
\beta$ is the ratio of Higgs vevs, and $R_f = \cot\beta\:(\tan\beta)$ for
weak isospin
+1/2 (-1/2) fermions.
Due to our choice of phases, there is a non-trivial phase associated with the
off-diagonal entries, which we denote by
$\mf(\overline{m}_{\ss f} e^{i \gamma_f})$, of the sfermion mass$^2$ matrix,
and
\beq
\overline{m}_{\ss f} e^{i \gamma_f} = R_f \mu + A^* = R_f |\mu|e^{i \theta_\mu}
+ |A|e^{-i \theta_A}
\eeq
 For a given value of $\mu$ and $A$, there are
then two phases which can be distinguished (by $R_f$), and we denote them by
$\gamma_t$ and $\gamma_b$.  The sfermion mass$^2$ matrix is diagonalized by
the unitary matrix $U$,
\beq
U = \pmatrix{ u & v \cr \noalign{\medskip} -v^* & u\cr }
\eeq
where $u^2 + |v|^2 = 1$, and we have taken $u$ real.

Previously \cite{fkmos} we considered the effect of sfermion mixing on
the relic density when the neutralino is mostly gaugino, and in
particular a bino.  The LSP is a bino whenever $M_2 < \mu$ and $M_2
\ga 200$ GeV; for smaller $M_2$, the bino is still the LSP for large
enough values of $\mu$.  The bino portion of the $M_2 - \mu$ LSP
parameter plane is attractive, as it offers the largest possibility
for a significant relic density \cite{osi3,gkt,mos}.  The
complementary portion of the parameter plane, with $\mu < M_2$, only
gives a sizable density in a limited region, due to the large
annihilation cross sections to $W^+W^-$ and $ZZ$ and due to
co-annihilations \cite{gs} with the next lightest neutralino (also a
Higgsino in this case), which is nearly degenerate with the LSP
\cite{my}.  We will also focus on binos as the LSP here.  In addition
to resulting in a sizable relic density, the analysis is
simplified by the fact that in the nearly pure bino region,
the composition and mass of the LSP is not very sensitive to the new
phases. However, as we will now show, the relic density, which is
determined primarily by annihilations mediated by sfermion exchange,
is quite sensitive to the phases, $\gamma_t$ and $\gamma_b$.

We begin by exploring the effect of the new phases on the relic
density of binos.  In the absence of these phases and in the absence
of sfermion mixing, there is an upper limit \cite{osi3,gkt} on the
bino mass of $\mb \la 250$ GeV for $\Omega h^2 < 1/4$. (This upper limit is
somewhat dependent on the value of of the top quark mass. In \cite{osi3,fkmos},
we found that the upper limit was $\sim 250 \gev$ for $m_t \sim 100 \gev$. When
$m_t = 174 \gev$, the upper limit is $260 \gev$. For
$m_t \sim 200 \gev$, this limit is increased to $\mb \la 300 \gev$.
Furthermore, one should be aware that there is an upward correction of about
15\% when three-body final states are included \cite{for} which raises the bino
mass limit to about $350 \gev$. This latter correction would apply to the
limits discussed below though it has not been included.)  As the bino mass is
increased, the sfermion masses, which must be larger than
$\mb$, are also increased, resulting in a smaller annihilation cross
section and thus a higher relic density. At $\mb \simeq 250$ GeV, even
when the sfermion masses are equal to the bino mass, $\Omega h^2 \sim
1/4$. Note that this provides an upper bound on the sfermion
masses as well, since the mass of the lightest sfermion is equal
to the mass of the bino, when the bino mass takes its maximum value.
When sfermion mixing is included \cite{fkmos}, the limits, which
now depend on the magnitude of the off-diagonal elements $m_f {\overline m_f}$,
are modified.  We find that this upper limit is relaxed considerably
when the phases are allowed to take non-zero values.

The dominant contribution to bino annihilation is due to sfermion
exchange and is derived from the bino-fermion-sfermion interaction
Lagrangian,
\begin{eqnarray}
{\cal L}_{f \tilde f \widetilde B } &=& {\textstyle{1\over\sqrt2}}g'
\left( Y_R \bar f P_L \widetilde B \tilde f_R + Y_L \bar f P_R \widetilde B
\tilde f_L \right) + \hbox{h.c.}   \nonumber \\
\noalign{\medskip}
&=& {\textstyle{1\over\sqrt2}}g' \bar f
    \left( \tilde f_1 x P_L + \tilde f_2 w P_L +
\tilde f_1 y P_R - \tilde f_2 z P_R \right) { \widetilde B } + \hbox{h.c.}
\end{eqnarray}
where $x = -Y_R\,v^*$, $y = Y_L\,u$,
$w = Y_R\,u$, $z = -Y_L\,v$, $P_{R,L}=(1 \pm
\gamma_5)/2$, $Y_R=2 Q_f$ and $Y_L=2 (Q_f-T_{3f})$ where $Q_f$ is the
fermion charge and $T_{3f}$ is the fermion weak isospin.

We compute the relic density by using the method described in ref.~\cite{swo}.
We expand
$\langle \sigma v_{ rel } \rangle$ in a Taylor expansion in powers of
$T/ m_{ \widetilde B }  $
\beq
\langle \sigma v_{ rel } \rangle = a + b\:({T/\mb}) + O\left(\left({T /
\mb}\right)^2\right)
\eeq
The coefficients $a$ and $b$ are given by
\beq
a = \sum_f v_f \tilde a_f
\eeq
\beq
b = \sum_f v_f \left[ \tilde b_f + \left( -3 + {3 m_f^2 \over 4 v_f^2
    m_{\widetilde B}^2 } \right) \tilde a_f \right]
\eeq
where $ \tilde a_f $ and $ \tilde b_f $ are computed from the expansion of the
matrix element squared in powers of $p$, the incoming bino momentum, and
$v_f = (1 - m_f^2/m_{\widetilde B}^2)^{1/2}$
is a factor from the phase space integrals.

We summarize the result by quoting the computed expression for $ \tilde a_f $:
\beq
\tilde a_f = { g'^4 \over { 128 \pi } }
  \left| {\Delta_1 ( m_f w^2 - 2 m_{ \widetilde B } w z + m_f z^2 ) +
  \Delta_2 ( m_f x^2 + 2 m_{ \widetilde B } x y + m_f y^2 ) \over
  \Delta_1 \Delta_2} \right|^2
\label{a}
\eeq
where $\Delta_i = m_{\tilde f_i}^2+m_{\widetilde B}^2-m_f^2$.
The result for $\tilde b_f$ is too lengthy for presentation here, but
was computed and used in the numerical integrations to obtain the
relic densities. The results reduce to the results quoted in
\cite{fkmos} in the limit of zero phases.

We show our results in Figures 1 and 2 for the upper and lower limits
on $\mb$ as a function of the magnitude of the off-diagonal term in
the top-squark mass$^2$ matrix, $\mtb$, given the conditions 1)
$\Omega_{\widetilde B} h^2 < 1/4$, $\,$ 2) the lightest sfermion is
heavier than the bino, and 3) the lightest sfermion is heavier than
$74\gev$\cite{alitti}.  In both figures we have taken $\tan \beta = 2$
and $m_{\rm top} = 174 \gev$. In Figure 1, $|\mu| = 3000$ GeV and in
Figure 2, $|\mu| = 1000$ GeV. The various curves are labeled by the
value of $\gamma_b$ assumed, and in addition, $\mb$ has been maximized
for all allowed values of $\theta_\mu$. The lower limit on $\mb$
assumes $\gamma_b = \pi/2$.  As one can see, when $\gamma_b$ is
allowed to take its maximal value of $\pi/2$, the upper limits are
greatly relaxed to $\mb \la 650$ GeV. With $|\mu|$ and $\gb$ fixed, for a given
value of $\mtb$ and $\thm$ all of the remaining quantities such as $|A|,
\theta_A, \gamma_t,$ and $\mbb$ are determined (though not necessarily
uniquely, as some are double valued).

The reason for the change in bounds becomes evident if we consider
$\Delta_1\approx\Delta_2\equiv\Delta$ and $\mf\ll\mb$ in (\ref{a}).  Then
\begin{eqnarray}
{\tilde a}_f \;&\approx&\; {{(g')}^4\over 8\pi\Delta^2}\: \mb^2\: Y_L^2\:
Y_R^2\:
u^2\;  {\rm Im}{(v)}^2 \;\;+ \;\;O(\mf\mb)\cr\noalign{\medskip}
      &=&\;\; {{(g')}^4\over 8\pi\Delta^2}\: \mb^2\: Y_L^2\: Y_R^2\: u^2\;
|v|^2\:\sin^2\!\gamma_{\ss f} \;\;+ \;\;O(\mf\mb)
\label{c}
\end{eqnarray}
The size of $u\,|v|$ depends on the quantity $r\equiv\mf\mfb/( \cos
2\beta(T_3 - 2 Q_f \sin^2\!\theta_W)M_Z^2)$; for $r \approx 1, \;
u\,|v| \approx .45$, while for $r\ll 1, u\,|v| \approx r$.  Note that
a non-zero value for $\sin \gamma_{\ss f}$ removes the p-wave suppression for
the
fermion $f$ (to a much greater extent than sfermion mixing alone
\cite{fkmos}) and greatly enhances the annihilation cross-section.  Of course,
for annihilations through the top quark channel, the p-wave suppression is not
terribly strong, as $m_t$ is not much smaller than $\mb$.

For large $\mtb$, the diagonal mass terms $M_L^2$ must be taken large
to ensure that the mass of the lightest stop is $\ga\mb$.  This drives
up the masses of the other sfermions and suppresses their contribution to the
annihilation.  As $\mtb$ is decreased, $M_L^2$ must drop, the other
sfermions begin to contribute and the upper
bound on $\mb$ is increased.  In particular, annihilation to $\mu$'s and
$\tau$'s becomes important, since
\beq Y_L^2 Y_R^2\,\Big|_{\,\mu,\tau}\; : \;
     Y_L^2 Y_R^2\,\Big|_{\,c, t}\; : \;
     Y_L^2 Y_R^2\,\Big|_{\,s, b} \; = \; 81\: : \: 4 \: : \: 1
\eeq
Decreasing $\gb$ reduces the effect of $\mu$'s and $\tau$'s, and this
can be seen as a decrease in the upper bounds in Figures 1 and 2.
For $\mtb$ sufficiently small, the stops become unmixed, diminishing
somewhat their contribution and slightly decreasing the upper bound
on $\mb$.

Lastly, we mention how the top contribution depends on $\gt$.
For the large values of $\mb$ allowed for $\mtb \la 1500\gev$, top
annihilation feels an enhancement from a non-zero $\sin\gt$ as in (\ref{c})
(but note that for the top, the higher order terms in $m_f$ are not
small).
However, except in a narrow region near $\mtb = 0$, the two stop
eigenstates $\tilde t_1$ and $\tilde t_2$
have a large mass splitting, and the bino couples with
different strengths to $\tilde t_1$ and $\tilde t_2$.  Taking $\gt$
away from $\pi$ decreases the coupling to the lighter stop (which can
offer a much greater contribution to the annihilation than can the
heavier stop), and we find that larger annihilation rates into tops
are generally found nearer to $\gt =\pi$ than $\gt = \pi/2$.  In
Figure 2, the sudden drop in the $\thm = \pi/8$ curve
 at $1500\gev$ occurs because
for $\mtb > (\tan\beta - \cot\beta)\,|\mu|\, , \gb - \pi/2 <
\gt < \gb + \pi/2$, and so for $\thm = \pi/8$, $\gt$ is prevented
from approaching $\pi$.

We turn now to the calculation of the electric dipole moments of the
neutron and the electron.  The EDM's of the electron and quarks
receive contributions from one-loop diagrams involving the exchange of
sfermions and either neutralinos, charginos, or (for the quarks)
gluinos. In the case of the neutron EDM, there are additional
 operators besides the quark electric dipole operator, $O_\gamma = {1 \over 4}
{\bar q} \sigma_{\mu \nu} q {\tilde F}^{\mu \nu}$
\cite{dn} which contribute. They are the gluonic operator
$O_G = - {1 \over 6} f^{abc}G_aG_b{\tilde G_c}$ \cite{w} and the quark color
dipole operator, $O_q = {1 \over 4}
{\bar q} \sigma_{\mu \nu} q T^a {\tilde G_a}^{\mu \nu}$ \cite{ads}. The gluonic
operator is the smallest \cite{other,ads} when all mass scales are taken to
be equal. These three
operators are conveniently compared to one another in \cite{aln} and
 relative to the
gluino exchange contribution to the $O_\gamma$ operator, it is found that
$O_\gamma : O_q : O_g = 21 : 4.5: 1$.

Because of the reduced importance of the additional operators contributing to
the neutron EDM, we will only include the three contributions to the quark
electric dipole moment. The necessary
$C\!P$ violation in these contributions comes from either
$\gf$ in the sfermion mass matrices or $\theta_\mu$ in the neutralino and
chargino mass matrices.  Full expressions for the chargino, neutralino and
gluino exchange contributions are found in \cite{ko}. The dependencies of the
various contributions on the CP violating phases can be neatly summarized:
for the chargino contribution
\beq
d_f^C \sim \sin\thm\;,
\eeq
with essentially no dependence on $\gf$; whilst for  the gluino contribution
\beq
d_f^G \sim \mfb \sin\gf\;,
\eeq
independent of $\theta_\mu$, and the neutralino contribution has pieces that
depend on both $\sin\thm$ and $\mfb\sin\gf$. All three contributions
can be important (including the neutralino contribution, in the case
of the electron EDM), and depending on $\sin\thm$ and $\sin\gf$, they
can come in with either the same or opposite signs.  In particular,
${\rm sign}[d_f^C/d_f^G] = {\rm sign}[\sin\thm/\sin\gf]$.  For the
mass ranges we consider, the dipole moments fall as the sfermion
masses are increased, and sfermion masses in the $\tev$ range can
bring these contributions to the
neutron and electron electric dipole moments below the
experimental bounds of $|d_n| < 1.1 \times10^{-25}e\:{\rm cm}$ \cite{nexp} and
$|d_e| < 1.9\times10^{-26}e\:{\rm cm}$ \cite{eexp},
even for large values of the $C\!P$ violating
phases \cite{ko}.  However, these large sfermion masses are inconsistent with
the cosmological bounds mentioned above, where sfermion masses must be
relatively close to the bino mass in order to keep the relic density in check.

We proceed as follows.  We fix the value of $\gb$ and take $|\mu| =
3000\gev$.  Then for several values of $\mtb$ between 0 and 1500
$\gev$, we determine the upper bound on $\mb$, as a function of
$\thm$.  As we vary $\thm$ across its full range, $\mbb$ and $\gt$
change, and this affects the annihilation rate (see (\ref{c}) and
following discussion), and consequently the bound on $\mb$.  Taking
$\mb$ at its maximum value allows us to take $M_L^2$ as large as
possible; although the electric dipole moments depend on $\mb$
as well, we find that the dependence on
$M_L^2$ is sufficiently strong that the EDM's take their minimum
values for the maximum values of $\mb$ and $M_L^2$.  We then compute
the quark and electron EDM's as a function of $\thm$ and $\mtb$,
and use the nonrelativistic quark model
to relate the neutron EDM to the up and down-quark EDM's via
\beq
d_n = {1\over3}(4d_d - d_u).
\eeq
If we find no region of the $\thm$--\,$\mtb$ parameter space which satisfies
both the neutron and electron EDM bounds, we decrease $\gb$ and repeat
the procedure.  In practice, we find the bound on the neutron EDM
the more difficult of the two to satisfy, and every region of the
parameter space we show which produces an acceptable neutron EDM also
produces a sufficiently small electron EDM.  We will consequently drop
further discussion of the electron EDM and concentrate on the neutron.

For the large value of $|\mu| = 3000\gev$, we find that the largest
contribution to the neutron EDM comes either from gluino exchange
(for the more negative values of $\thm$) or chargino exchange (for the
more positive values of $\thm$),
and that the value for $|d_n|$ is too large unless $\gb$ takes a relatively
small value.  In particular, we find non-negligible experimentally
acceptable regions of the parameter space only for $\gb \la \pi/25$.
In Figure (3) we show a contour plot of the neutron EDM as a function
of $\thm$ and $\mtb$ for $\gb = \pi/40$.  The shaded regions demarcate
the range of $\thm$ for this choice of $\gb$.  Much of this range
produces a sufficiently small $|d_n|$.  As we increase $\mtb$, the
$\tilde d$ and $\tilde u$ masses become large and $|d_n|$ falls.  As we
move to values of $\mtb$ greater than $\sim 1500\gev$, we begin to
require a significant tuning of $M_L^2$ to produce
$\Omega_{\widetilde B} h^2 < 1/4$.

Near the boundaries of the allowed range of $\thm$, $\gt$ approaches
$\gb\pm\pi/2$.  As we explain above, the top contribution to the bino
annihilation drops off as we move away from $\gt=\pi$, and the upper
bound on the bino mass, and sfermion masses, falls.  It is these lower
values for the sfermion masses which are primarily responsible for the
sharp rise in $|d_n|$ near the boundaries.  There is one last subtlety
which requires mention.  For every set of the parameters \{$|\mu|,
\mtb, \gb, \thm$\}, there are two possible values for \{$\gt, \mbb$\}.
We find that one of the two sets of values always gives a smaller
$|d_n|$, given the cosmological constraints on the bino and sfermion
masses, and it is these smaller values of $|d_n|$ which we plot.

We repeat this procedure for $|\mu| = 1000\gev$.  For lower $|\mu|$,
the chargino exchange contribution is enhanced relative to the gluino
exchange contribution.  In Figure (4), we show a contour plot of $d_n$
for $\gb = \pi/8$.  At ($\mtb= 1500\gev,\, \thm = 0)\,, d_n$ vanishes,
as $\mbb = \thm = \sin\gt = 0$.  Of course the gluonic and quark color
dipole contributions to $d_n$ will not vanish everywhere along the
contour $d_n = 0$, but their contributions are $\la 3\times 10^{-26}
e\:{\rm cm}$ in the region plotted.  Also, the gluino and chargino
contributions scale differently as $\mb$ and $\msf$ are changed.  If
we take $\mb$ and $\msf$ less than their maximum values in Figure (2),
the contours of Figure (4) will shift, by $\la 1.\times 10^{-25}$.
One should therefore concentrate on the qualitative features of
 Figure (4), as the exact positions of the contours are not significant.

In summary, we have found that $C\!P$ violating phases in the MSSM can
significantly affect the cosmological upper bound on the mass of an
LSP bino.  In particular, taking the maximal value $\pi/2$ for the
phase $\gb$ of the off-diagonal component of the $T_3 = -1/2$ sfermion
mass matrices pushes the upper bound on $\mb$ up from $\sim 250\gev$
to $\sim 650\gev$.  When we additionally consider constraints on
neutron and electron electric dipole moments, we find the upper bound
on $\mb$ is reduced to $\sim 350\gev$.  Various combinations of the
$C\!P$ violating phases are constrained as well: $|\thm|\la 0.3$ and
$|\gb|\la\pi/6$ for $|\mu| \ga 1000 \gev$, while $\gt$ and $\theta_A$
are essentially unconstrained.
We note that although the bounds on $\thm$ and $\gb$ are small, they
are much larger than the values of order $10^{-3}$ typically considered.

\vskip 1in
\vbox{
\noindent{ {\bf Acknowledgments} } \\
\noindent  This work was supported in part by DOE grant DE--FG02--94ER--40823
and NSF grant PHY--91--16964.}


\vskip 2in
\noindent{\bf{Figure Caption}}

\vskip.3truein

\begin{itemize}
 \item[]
\begin{enumerate}
\item[]
\begin{enumerate}

\item[Fig.~1)]Upper limits on the bino mass as a function of the off-
diagonal element $\mtb$ in the top squark mass$^2$ matrix, for various values
of
$ \gamma_b$, the argument of the off-diagonal element of the $T_3 = -1/2$
sfermion mass$^2$ matrix. Also shown is the lower bound (lowest curve) on the
bino mass assuming
$\gamma_b =
\pi/2$. The value of $|\mu|$ was chosen to be $3000 \gev$.

\item[Fig.~2)] As in Fig. 1, with $|\mu| = 1000 \gev$.

\item[Fig.~3)] Contours of the neutron electric dipole moment, $d_n$, in the
$\thm$ -- $\mtb$ plane in units of $10^{-25}e\:{\rm cm}$. The value of $|\mu|$
was chosen to be $3000 \gev$. The shaded region corresponds to values of $\thm$
and $\mtb$ which are not allowed algebraically for this value of $\mu$ and
$\gb$.

\item[Fig.~4)] As in Fig. 3, with $|\mu| = 1000 \gev$.
\end{enumerate}
\end{enumerate}
\end{itemize}
\newpage

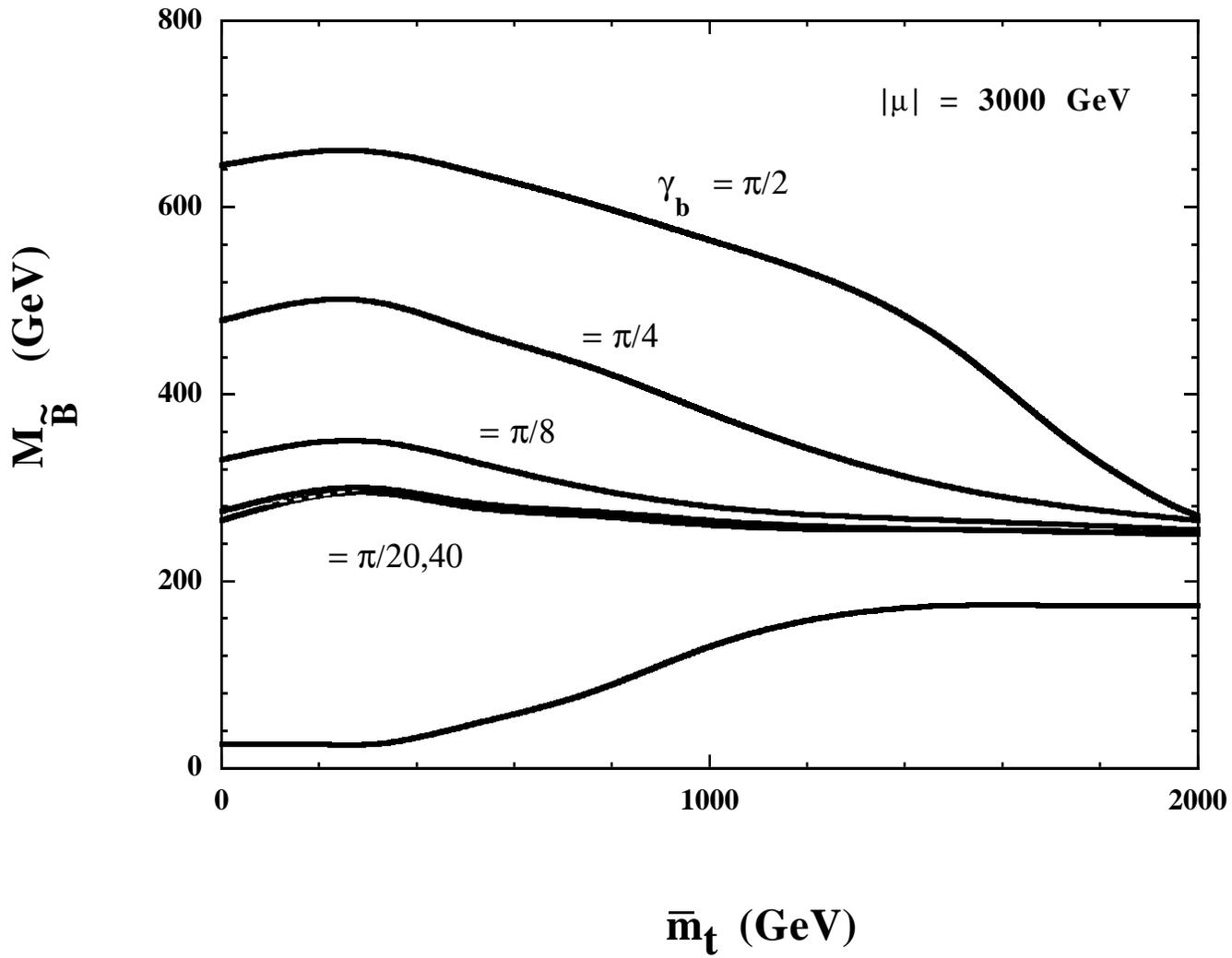
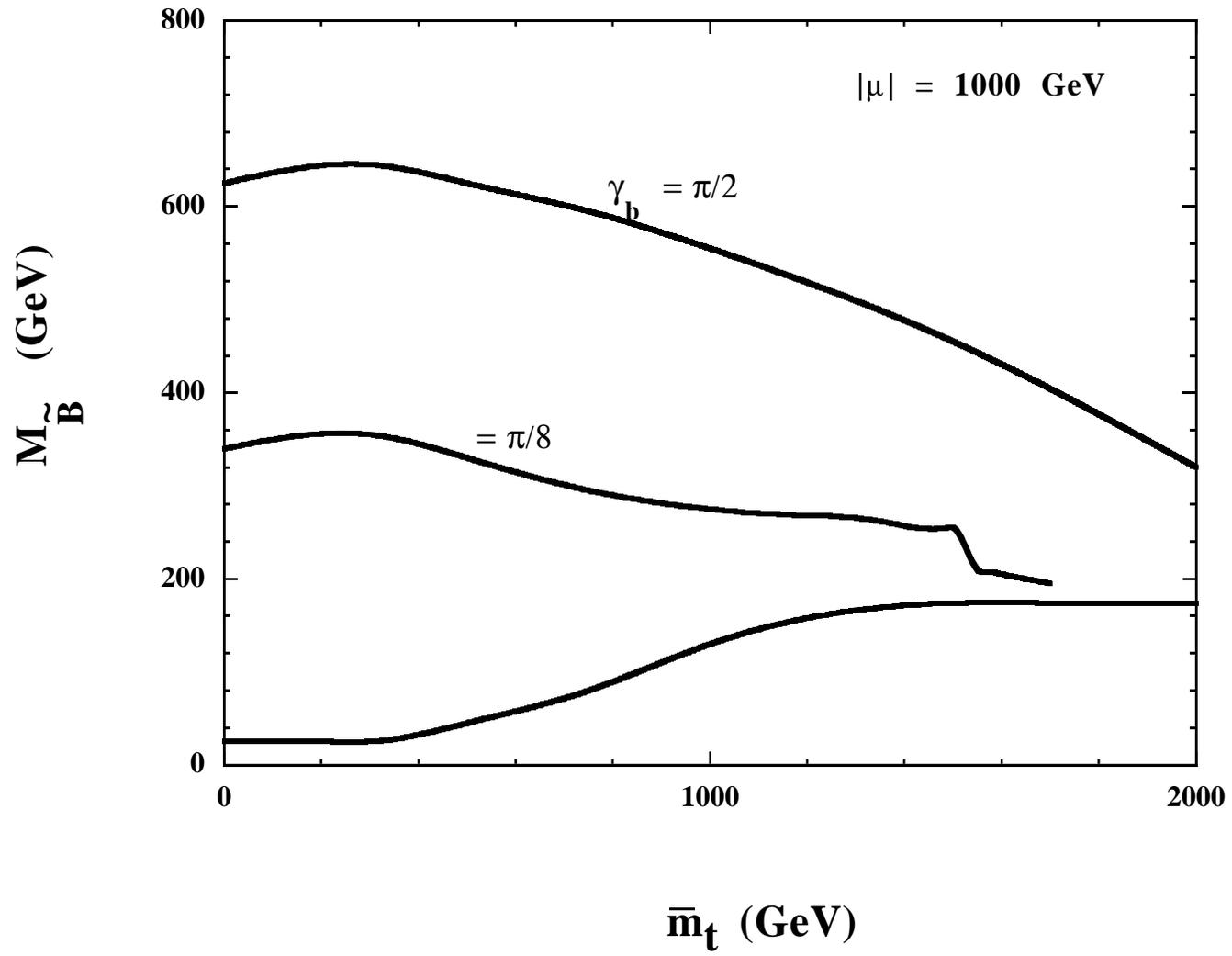
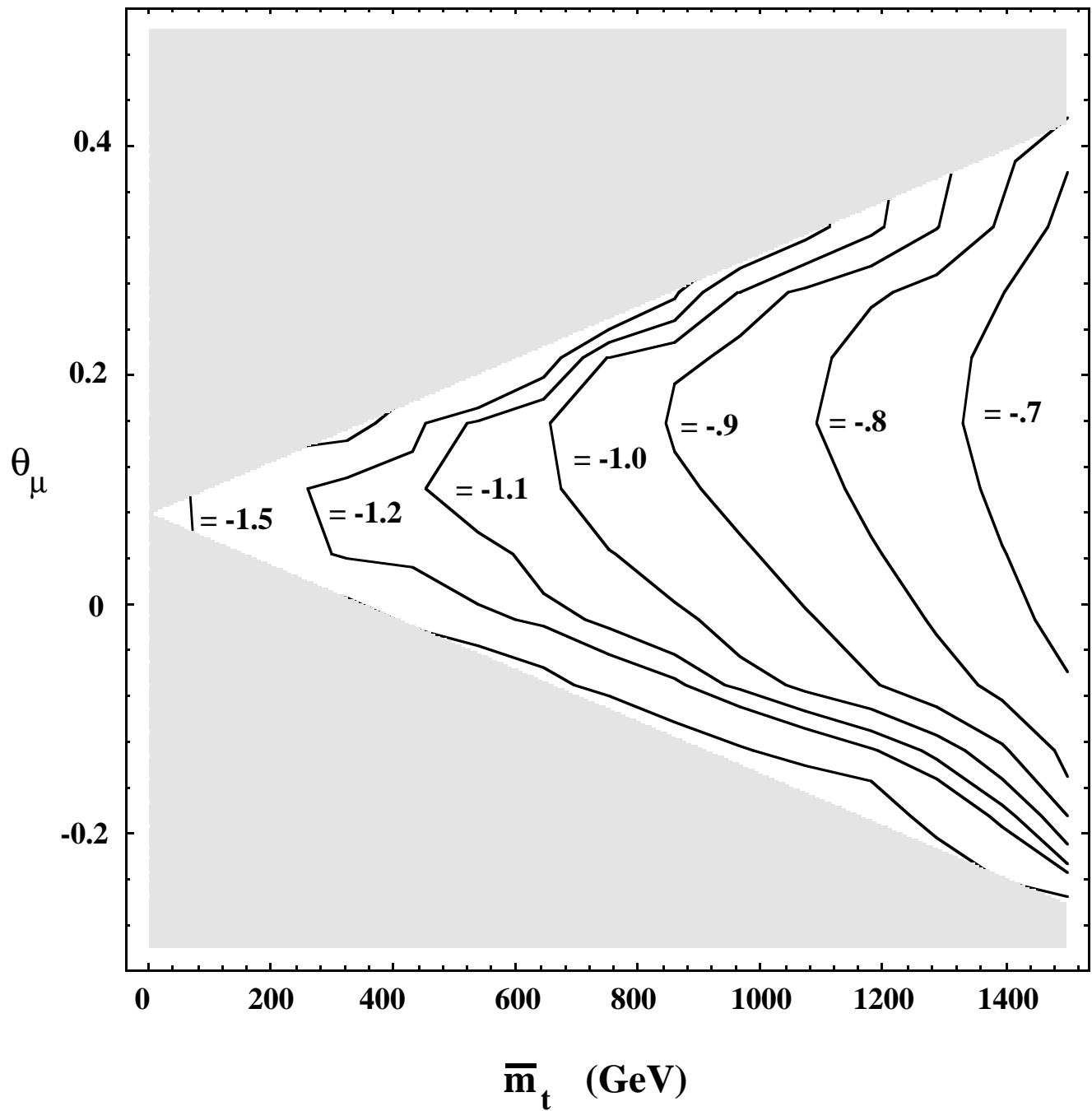
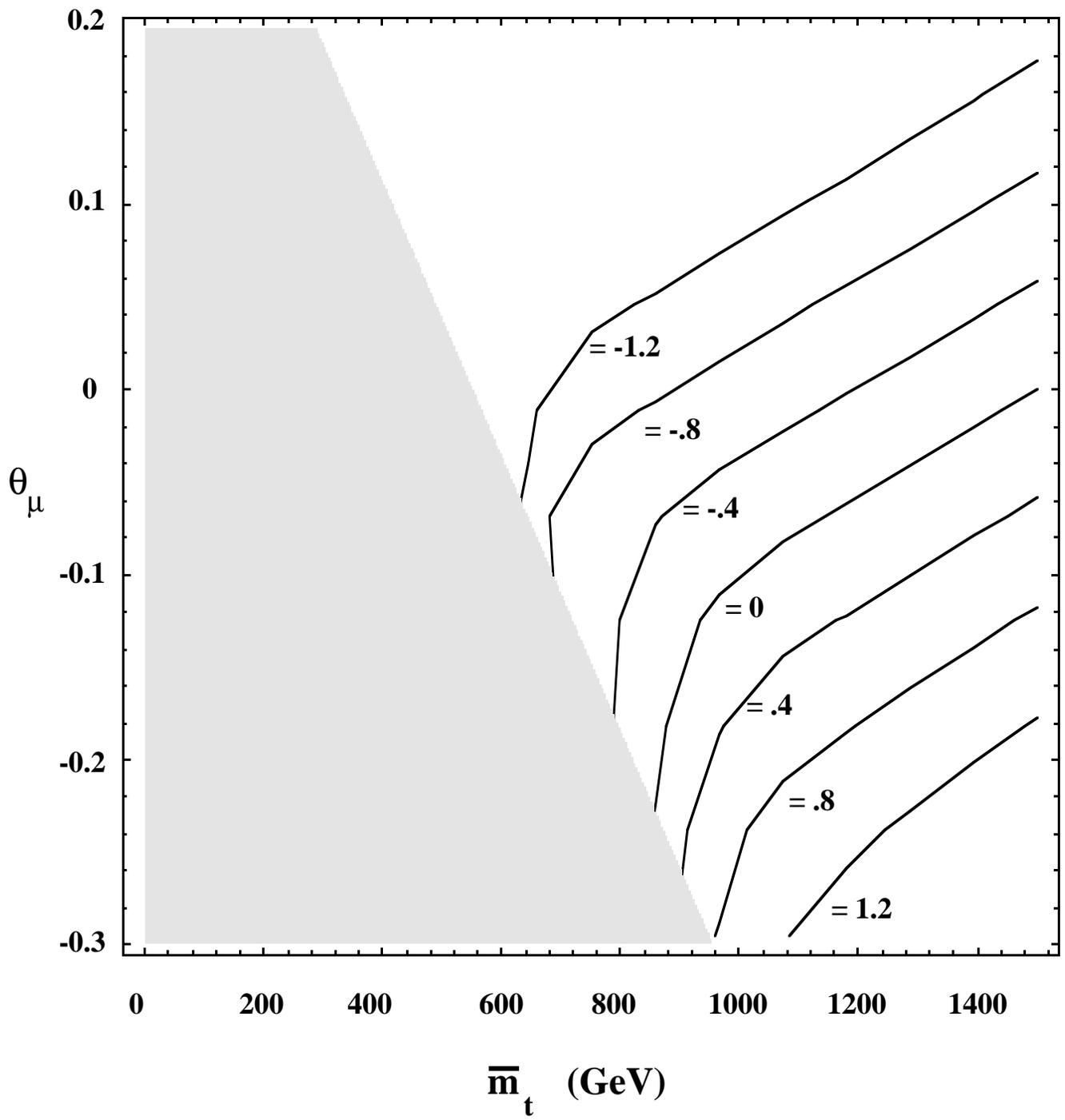
\end{document}